**From Verification to Amplification: Auditing Reverse Image Search as Algorithmic**

**Gatekeeping in Visual Misinformation Fact-checking**

Cong Lin[1], Yifei Chen[2], Jiangyue Chen[3], Yingdan Lu[4] , Yilang Peng[5], Cuihua Shen[6]

[1] School of Journalism and Communication, Tsinghua University, Beijing, China.

[2] Department of Public Management and Policy, Georgia State University, Atlanta, Georgia, USA.

[3] School of Journalism and Communication, The Chinese University of Hong Kong, Hong Kong.

[4] Department of Communication Studies, School of Communication, Northwestern University, Evanston, USA.

[5] Department of Financial Planning, Housing and Consumer Economics at the University of Georgia, Athens, Georgia, USA.

[6] Department of Communication, University of California, Davis, Davis, California, USA.

[*] Corresponding Author:

Cuihua Shen, Address: One Shields Ave, Davis, California, 95618, USA. Email: cuishen@ucdavis.edu

**Declaration of conflicting interests**

**Funding**

The author(s) received no financial support for the research, authorship, and/or publication of this article.

**Data availability statement**

Data and code are available at

https://osf.io/f8us9/overview?view_only=f1ee786a1a7b4bac875f357c6a1b72f7.

**ORCID IDs**

Cong Lin https://orcid.org/0009-0000-8546-8021

Yifei Chen https://orcid.org/0009-0006-4630-6162

Jiangyue Chen https://orcid.org/0009-0001-8949-4603

Yingdan Lu http://orcid.org/0000-0002-9955-6070

Yilang Peng http://orcid.org/0000-0001-7711-9518

Cuihua Shen http://orcid.org/0000-0003-1645-8211

**Author Biographies**

**Cong Lin** is a Ph.D. Student in the School of Journalism and Communication,Tsinghua University. His research interests include computational social science, visual communication, human-computer interaction and algorithm studies.

**Yifei Chen** is a Ph.D. Student in the Andrew Young School of Policy Studies at Georgia State University. Her research interests include technology and public policy, computational social science, and local economic development.

**Jiangyue Chen** is a Ph.D. Student in the School of Journalism and Communication at The Chinese University of Hong Kong. Her work focuses on algorithm studies, social media platforms, and computational social science.

**Yingdan Lu** (Ph.D., Department of Communication, Stanford University) is an Assistant Professor in the Department of Communication Studies, Northwestern University. She uses computational and qualitative methods to understand topics like the evolution and engagement of digital propaganda in authoritarian regimes and how individuals encounter and communicate multimodal information in AI-mediated environments.. For more information, see her website: https://yingdanlu.com/.

**Yilang Peng** (Ph.D., Annenberg School for Communication, University of Pennsylvania) is an Associate Professor in the Department of Financial Planning, Housing and Consumer Economics at the University of Georgia. His scholarship is at the intersection of computational social science, visual communication, science communication, and social media analytics.

**Cuihua Shen** (Ph.D., University of Southern California) is a Professor at the Department of Communication, University of California, Daivs. Her research interests include online networks, misinformation, and computational social science.

**Abstract**

As visual misinformation becomes increasingly prevalent, platform algorithms act as intermediaries that curate information for users' verification practices. Yet, it remains unclear how algorithmic gatekeeping tools, such as reverse image search (RIS), shape users' information exposure during fact-checking. This study systematically audits Google RIS by reversely searching newly identified misleading images over a 15-day window and analyzing 34,486 collected top-ranked search results. We find that Google RIS returns a substantial volume of irrelevant information and repeated misinformation, whereas debunking content constitutes less than 30% of search results. Debunking content faces visibility challenges in rankings amid repeated misinformation and irrelevant information. Our findings also indicate an inverted U-shaped curve of RIS results page quality over time, likely due to search engine "data voids" when visual falsehoods first appear. These findings contribute to scholarship of visual misinformation verification, and extend algorithmic gatekeeping research to the visual domain.

## Introduction

In the contemporary media landscape where information overload is common and users' attention is limited (Lanham, 2006), digital platforms have the power to shape the visibility of online content and users' information exposure (Bucher, 2012; Wallace, 2018). In particular, as online misinformation becomes increasingly prevalent (Vosoughi et al., 2018), platform algorithms act as gatekeepers that directly modulate users' exposure not only to misinformation flows but also to fact checks that counteract such misinformation (Aslett et al., 2023; Chuai et al., 2024). For instance, social media platforms can algorithmically moderate misleading content (Gillespie, 2022) and users often rely on search engines to verify the veracity of information they encounter elsewhere (Hasanian and Elsayed, 2021). While algorithms could help users access debunking messages for information verification, they may also inadvertently amplify misinformation by prioritizing popular, yet misleading, content (Fernandez et al., 2024).

Nevertheless, with only a few exceptions (Aslett et al., 2023; Pröllochs, 2022), the role of platform algorithms in misinformation verification remains underexplored in the fact-checking literature. Most existing studies focus on microlevel factors that shape users' perceptions of information credibility, such as linguistic features of fact-checking labels (Xue et al., 2024), source credibility of fact checks (Liu et al., 2025), and the presentation order of debunking content (Peter and Koch, 2016), rather than the platform mechanisms that determine what content is surfaced during verification. Among all modalities of misinformation, visual misinformation has become increasingly pervasive, posing challenges for fact-checking. For example, images are susceptible to manipulation and easily repurposed across contexts to create misleading content (Heley et al., 2022). Understanding how algorithms process visual content is thus critical for navigating the multimodal media environment. However, prior studies on algorithmic gatekeeping focus predominantly on textual information (Møller, 2022; Yang and Peng, 2022). Expanding this line of inquiry to visual content could help reveal how algorithms influence access to visual content and address the unique challenges of verifying visual misinformation. In contemporary information ecosystems, reverse image search (RIS) has emerged as one of the primary interventions for verifying visual content (Qian et al., 2023). For example, Google RIS has been widely employed for countering visual misinformation (Juneja and Mitra, 2022). Yet, it remains unclear how RIS algorithms curate and rank the image results during verification processes, and how misinformation features further complicate algorithmic gatekeeping.

To address these gaps, this study leverages an auditing approach to examine how Google RIS shapes users' information exposure in the verification process. We collected 95 misleading images from fact-checking organizations, and conducted seven rounds of reverse image searches for each within a 15-day window, which generated 34,486 top search results and associated metadata including sources, links, and snapshots. We assessed RIS performance by analyzing information relevance, veracity, and rankings of search results, while also investigating how RIS results page quality varied across misinformation types, topics, and search timing.

Our findings make several contributions. Theoretically, this study discusses how algorithms curate information feeds for users in misinformation verification processes, and extends understandings of algorithmic gatekeeping to visual content, demonstrating that RIS algorithms could surface both debunking content and falsehoods. Methodologically, we develop a framework that integrates dynamic algorithm auditing with content analysis, which can be usefully applied to examine platform-mediated fact-checking. Practically, we find that "data voids" exist and RIS can amplify visual misinformation, underscoring the need for caution when verifying emerging misleading images and highlighting the importance of platform accountability and users' digital literacy.

## Theories and Research Questions

### Algorithmic Gatekeeping and Misinformation Amplification

Gatekeeping theory in journalistic contexts traditionally posits that, as gatekeepers, editors in newsrooms and media organizations can select and shape which messages or news stories would be covered by mass media (Shoemaker and Vos, 2009; Wallace, 2018). More recently, scholars have identified emerging actors playing gatekeeping roles, such as platform algorithms (Wallace, 2018; Yang and Peng, 2022). For instance, users' news consumption experiences are increasingly mediated by news aggregators (Møller, 2022), and search engines algorithmically retrieve and rank content relevant to users' queries (Van Couvering, 2007). Unlike traditional gatekeeping grounded in journalistic norms and news values (Wallace, 2018), algorithms exert power through information visibility and public attention allocation, as they can decide selectively what content to show, filter out, emphasize, or diminish (Bucher, 2012; Diakopoulos, 2019; Yang and Peng, 2022), shaping the composition and structure of users' information feeds (Thorson and Wells, 2016).

As misinformation becomes increasingly prevalent, platform algorithms also play an important role in modulating the exposure to fake news and harmful content (Gillespie, 2022). Broadly, misinformation describes false or misleading information regardless of the motivation of the falsehood (Lazer et al., 2018). Acting as gatekeepers, social media algorithms can help flag misleading content to raise users' awareness and attention (Chuai et al., 2024; Pröllochs, 2022). Also, search engine algorithms may retrieve relevant and useful information for claim verification (Hasanian and Elsayed, 2021).

However, algorithms differ from journalists in their gatekeeping across the stages of information access, selection, and publication, often inadvertently amplifying misinformation (Wallace, 2018). Search engine algorithms, for example, rely on a pool of structured data and automated processing, potentially incorporating unreliable sources and misleading content, which traditional journalistic gatekeepers would likely exclude. Algorithms follow mathematical logic rather than normative values such as content quality, thereby excluding criteria that cannot be operationalized algorithmically (Wallace, 2018). For example, objectivity and fairness are core elements of journalistic assessments of media quality, but they are not key determinants in search algorithms as they are difficult to quantify (Van Couvering, 2007). Finally, algorithmic gatekeepers such as search engines and news aggregators publish content on their proprietary platforms (Wallace, 2018). As market-driven actors seeking to promote user engagement and time spent, these platforms have incentives to orient the recommendation results in self-serving ways, which increases the likelihood of surfacing biased, manipulated, or deceptive content (Diaz Ruiz, 2025; Rieder and Sire, 2014).

While the potential of algorithmic amplification of misinformation is concerning, limited research has bridged algorithmic gatekeeping and misinformation verification scholarship to examine how algorithms curate information flows during users' verification practices. This line of inquiry is crucial, as the content users encounter likely shapes their fact-checking outcomes and beliefs, which in turn shape their subsequent information-sharing behaviors.

**Search Engines as Algorithmic Gatekeepers in Misinformation Verification**

Faced with the proliferation of online misinformation, professional news outlets and fact-checking agencies act as gatekeepers by verifying claims and publishing correction posts, while platform algorithms moderate false content or reduce its visibility (Gillespie, 2022; Juneja and Mitra, 2022). In most cases, individuals encounter such debunking messages incidentally, as they consume news or browse social media.

Meanwhile, individual users can also take a more active role in verification, such as querying search engines to assess information credibility. Unlike passive exposure to corrective content, users' proactive behaviors reflect more deliberate and participatory forms of verification (Matanji et al., 2024; Qian et al., 2023). However, without professional guidance, individuals tend to follow their personal interests and predispositions (Wallace, 2018), rather than established fact-checking principles such as scientific evidence or source credibility. In this context, it becomes crucial to investigate how search engines, as algorithmic gatekeepers, shape the information users encounter during user-initiated fact-checking processes.

Notably, a growing body of research has raised critical concerns about quality of search engine results and the effects on misinformation verification. For example, search engine algorithms could return both accurate and misleading content, surfacing or amplifying junk news (Bradshaw, 2019; Hasanian and Elsayed, 2021). Also, the ranking of search results can be manipulated by search engine optimization (SEO) strategies (Rieder and Sire, 2014). For example, by purchasing keywords, fake news websites can secure higher placement in search results (Bradshaw, 2019), diverting users' attention from credible information sources. In this regard, users' self-initiated fact-checking via search engines may struggle to achieve corrective outcomes. Aslett et al. (2024) demonstrate that searching online to verify misinformation may paradoxically increase users' perceived veracity of misinformation. One explanation is that in contexts where reliable content is sparse, users risk entering "data voids" where search engines return little credible content and place non-credible information at the top of results (Boyd and Golebiewski, 2020). Such outcomes are not only enabled by SEO practices but also by the recycling of content across low-quality publishers, creating misinformation ecosystems that pollute search results (Aslett et al., 2024). In summary, misinformation verification outcomes are not only determined by platform interventions (e.g., content moderation) or fact-checkers' behaviors (e.g., posting verified messages), but also conditioned by algorithmic infrastructures.

**Reverse Image Search as Algorithmic Gatekeeping**

Visual misinformation, which refers to visual content containing false and inaccurate presentations of information, has become increasingly prevalent in the digital media environment (Peng et al., 2023; Yang et al., 2023). Compared to text, visuals capture greater attention, enhance emotional arousal, can be recalled more efficiently, and are often perceived as more credible due to the indexical qualities of images that reinforce realism heuristics (Hameleers et al., 2020; Weikmann and Lecheler, 2023). RIS is a widely adopted tool for verifying visual misinformation (Juneja and Mitra, 2022), which allows users to submit images as queries to search the web and returns the matched visuals (Qian et al., 2023). While previous gatekeeping research has primarily focused on textual content such as news stories or social media posts (Møller, 2022; Yang and Peng, 2022), little is known about how visual search reshapes algorithmic gatekeeping.

Notably, search-by-image algorithms may operate through distinct logics from text-based ones. First, while text-based search and recommendation are driven by semantic or lexical matching to assess topical alignment, visual search systems like RIS rely on perceptual similarity involving pixel-level features of image content rather than metadata such as keywords or descriptions (Gaillard and Egyed-Zsigmond, 2017; Hao et al., 2021). While these algorithms can help users find the same image across different contexts, thereby providing cues for assisting image credibility assessment (Matanji et al., 2024), perceptual similarity often diverges from factual or contextual relevance. As a result, users may retrieve pictorially similar but out-of-context images with misleading claims, increasing the risk of exposure to visual misinformation. Second, the complexity of images poses greater challenges for algorithmic processing than text. Images are particularly vulnerable to manipulation or editing, often leaving no visible trace (Weikmann and Lecheler, 2023). Traditional gatekeeping demands that journalists act as "gatecheckers" to manually verify image authenticity before publication, even though this process is highly time- and effort-consuming (Schwalbe et al., 2015). Given the lack of technical tools in detecting manipulated and synthetic visuals (Lu et al., 2023), algorithms are unequipped to verify image authenticity like traditional journalists and may retrieve visual misinformation, particularly when users submit manipulated or synthetic images as queries.

**Algorithmically Curated RIS Results**

Algorithmic gatekeeping operates through the allocation of public attention, shaping both which information users encounter and the ranking in which it appears on digital platforms (Yang and Peng, 2022). Information **relevance** serves as a key indicator of search quality (Van Couvering, 2007), as users with limited attention can be distracted by irrelevant results that waste time and reduce the efficiency of information acquisition. However, relevance alone does not ensure accuracy, especially in contexts where misinformation circulates widely. The **veracity** of retrieved information is another crucial criterion (Aslett et al., 2024). Together, relevance and veracity represent two complementary dimensions of search performance, namely, whether information is both contextually relevant and factually correct.

Another critical aspect of algorithmic gatekeeping is information visibility through the **ranking** of search results (Bucher, 2012). Research consistently shows that users focus disproportionately on top-ranked items (Lorigo et al., 2008), making ranking logics central to how search engines curate results. In the context of RIS algorithms that retrieve visually similar images and accompanying information from the web, it is important to examine how these ranking mechanisms operate in a multimodal environment, and whether RIS algorithms prioritize credible or misleading content when users reverse search misleading visuals. Therefore, we propose the following research questions:

**RQ1: When using RIS for visual misinformation verification, how are search results distributed in terms of (a) information relevance and (b) information veracity?**

**RQ2: When using RIS for visual misinformation verification, how are search results with varied information relevance and veracity ranked on results pages?**

At the results page level, the above three indicators-information relevance, veracity and ranking-contribute to the overall quality of RIS results pages (Hussein et al., 2020; Van Couvering, 2007). Here, "results page quality" represents a comprehensive approach to assessing RIS gatekeeping, integrating all search results to understand how effectively the algorithm supports reliable verification outcomes.

Drawing on prior studies (Heley et al., 2022; Thomson et al., 2024), visual misinformation, specifically images, can be divided into three categories: *out-of-context images* are authentic images that have been placed together with inaccurate captions; *manipulated images* are visual content that has been edited in certain ways to distort interpretation; *AI-* *generated images* are those wholly fabricated by AI tools. Each type poses distinct computational challenges as RIS algorithms try to find visually similar content. Compared to manipulated images, out-of-context visual misinformation challenges RIS algorithms because, despite visual matching, RIS alone does not assess the alignment between images and accompanying captions (Gaillard and Egyed-Zsigmond, 2017), potentially yielding irrelevant or misleading results for verification. Also, compared to those of manipulated and out-of-context misinformation, queries involving AI-generated visuals may yield lower-quality results because such content often lacks referential matches in training or cross-validation datasets, limiting the algorithm's ability to establish reliable context (Lu et al., 2023). Together, we anticipate that RIS results page quality would vary across visual misinformation types, as each type affords different levels of algorithmic processing capacity and data availability.

Moreover, the topics of visual misinformation, particularly political versus non-political ones, may receive unequal debunking attention from media and fact-checking institutions, resulting in variations in RIS results page quality. Compared to non-political topics such as entertainment-related misinformation, political or election-related misinformation raises civic and democratic concerns and is more directly tied to the public interest (Mosleh et al., 2024). Professional news organizations and fact-checkers often prioritize the verification and correction of politically salient claims and respond more rapidly to such content (Cazzamatta, 2025). Earlier and more concentrated availability of debunking content increases the likelihood that RIS algorithms retrieve corrective information when users query related images, potentially yielding higher quality RIS results pages for political visual misinformation.

Additionally, a persistent time lag often exists between the emergence of misinformation and the publication of corresponding fact checks (Pilarski et al., 2024). This highlights the importance of temporality in misinformation verification. Indeed, algorithmic gatekeeping differs from traditional editorial practices in its temporal dynamics. Unlike the manual and episodic editorial decision-making typical of newsrooms, algorithmic curation is a dynamic and continuous process. Content visibility is constantly recalibrated as algorithms re-rank information flows in response to changing engagement metrics such as clicks and reposts on platforms (Bucher, 2012). This temporal dynamic is salient in the visual misinformation verification context, where visibility may shift as the volume and popularity of misinformation and debunking content evolve over time (Kauk et al., 2025). Accordingly, the quality of RIS results pages should be conceptualized not as static but as temporally contingent, varying with shifts in both user engagement and content availability. Taken together, these technical, normative, and temporal considerations suggest that RIS performance is multidimensional and dynamic. Therefore, we ask:

**RQ3: How does the quality of RIS results page vary across (a) visual misinformation type, (b) political versus non-political visual misinformation, and (c) search timing?**

## Method

### Data Collection

We conduct an audit study to collect and analyze a large-scale, annotated dataset of Google RIS results. Google is the most popular search engine worldwide[1] and its RIS tool has been widely employed to counter visual misinformation (Juneja and Mitra, 2022). Also, algorithm audits can provide a systematic appraisal of potential problems in socio-technical systems through iterative input queries and observations of algorithmic outputs (Metaxa et al., 2021). The data collection processes are presented in Figure 1.[2]

**Figure 1**

*Data Collection Processes.*

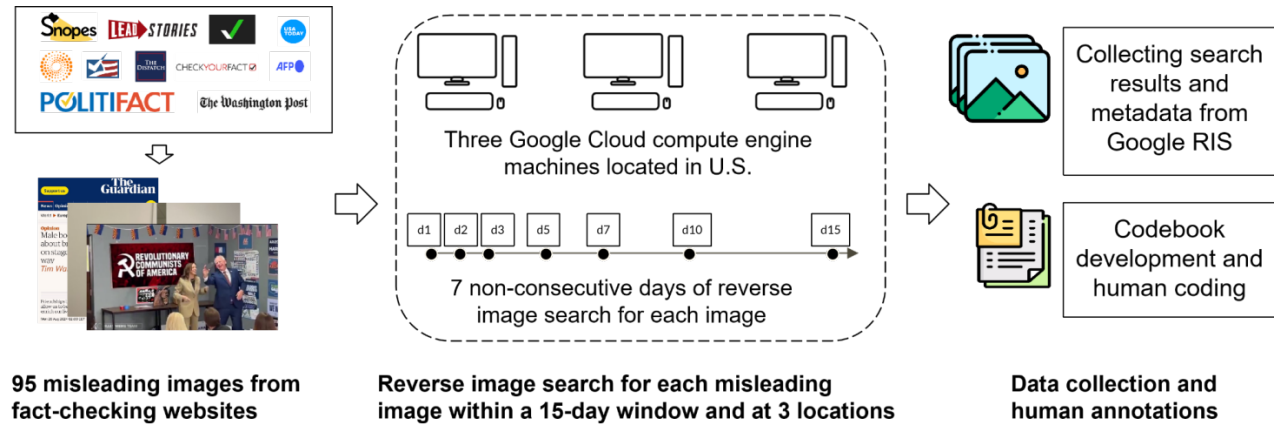

**Collecting Visual Misinformation**

We first collected visual misinformation from professional fact-checking websites to obtain recently-verified misleading images. We selected eleven active and U.S.-based fact-checking websites that were frequently cited in previous literature (e.g., Liu et al., 2025; Singer, 2023; see Appendix A.1 for the full list). The misleading images were collected over seven days across a two-week period (August 19 to September 1, 2024): Monday, Wednesday, and Friday of the first week, and Tuesday, Thursday, Saturday, and Sunday of the second week. This design reduces potential bias that could arise from single-week or consecutive-day sampling. On each day of data collection, we visited all eleven fact-checking websites from 10 p.m. to 12 a.m. (U.S. Eastern Time) to identify fact checks published on that day (d0) that debunked misleading images. Misleading images include manipulated images, AI-generated images, and unedited images placed out of context (see Appendix A.2 for criteria and examples). In total, we collected 105 misleading images. After removing duplicates and low-resolution images (e.g., screenshots from misleading videos), 95 misleading images remained.

**Figure 2**

*Illustrated Glossary of Study Terms.*

**1** RIS results page

RIS returns two types of results pages: "Visual Matches" page which presents search results in a grid layout, and "Exact Matches" which presents search results vertically in one column.

**2** Search result

A search result includes the image, accompanying text captions, source, link, and rank on the RIS results page.

"Visual Matches" page

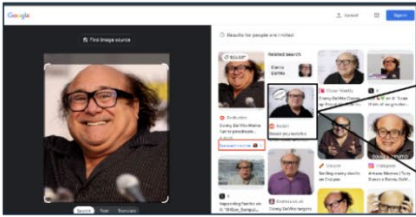

"Exact Matches" page

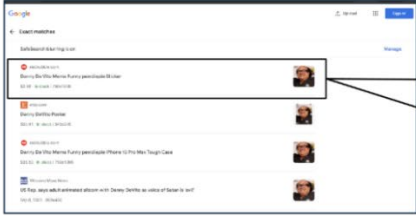

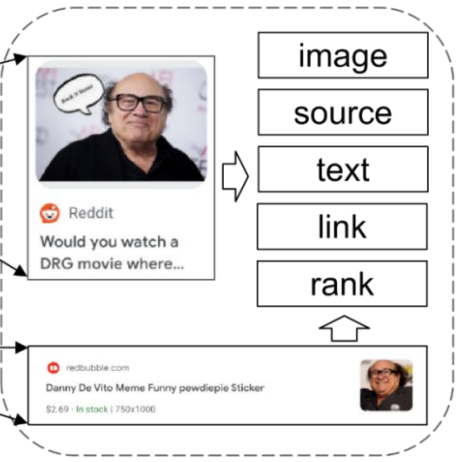

image

source

text

link

rank

**3** Search result link

The directed URL of a search result. A link can appear across results pages, search time, and misleading images.

**4** Snapshot of search result links

A snapshot of the destination webpage of a search result link.

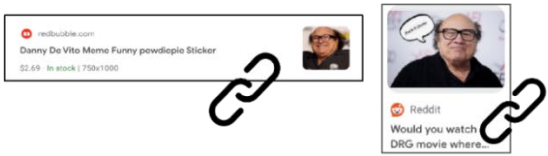

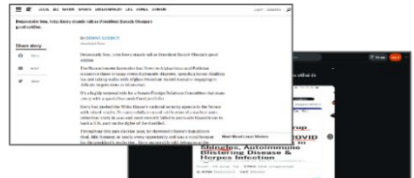

**5** A misleading image–search result link pair

A pair consists of a misleading image and one corresponding search result link, which is annotated using the fact-checking article associated with that image.

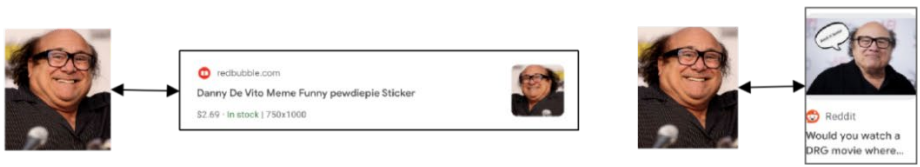

### Performing Reverse Image Search

We performed RIS on every misleading image automatically using Google Cloud compute engine machines. We set up three virtual machines with the same architecture, configuration, and operating system (64bits, Debian 12, 2 GB RAM), respectively located in the U.S. Western, Central, and Eastern regions. This design seeks to control machine-related variability while also maintaining geolocation diversity. The machines were logged out of Google accounts and cookies were cleared to minimize personalization. For each misleading image, the first search happened on the next day (d1) after the fact-checking article was published (d0). Subsequent searches were repeated on the 2nd (d2), 3rd (d3), 5th (d5), 7th (d7), 10th (d10), and 15th (d15) days. This temporal pattern is informed by prior work indicating that online collective attention frequently follows log-normal distributions, characterized by rapid early decay (Wu and Huberman, 2007). Even on platforms with relatively slower attention decay, such as YouTube, estimated content half-life falls within 8 days (Graffius, 2022), suggesting that a 15-day window is sufficient to capture concentrated attention before it wanes.

Google RIS yields two types of results pages: the "Visual Matches" page and the "Exact Matches" page (see Figure 2). The former displays image results with similar visual features, while the latter returns images that are identical or nearly identical to the image submitted.[3] On both pages, our machines automatically scraped metadata for each search result, including title text, source name, link, and its display position.

### Snapshots of Search Result Links

We manually downloaded the snapshots of unique search result links as local HTML files. Previous eye-tracking research indicates that most users tend not to go beyond the first page when Googling (Lorigo et al., 2008), so we focus on the first pages of RIS only. On "Visual Matches" pages, results are presented in a grid of multiple rows and three columns,[4] and only the first four rows were downloaded (all on the first page; three results within each row and 12 search results at maximum). On "Exact Matches" pages, search results are displayed in a single vertical column, and up to the top 10 results (all on the first page before loading more) were downloaded. The same search result link could appear repeatedly across RIS results pages, search timings, and misleading images. For these links, we only downloaded their snapshots once (see Appendix A.3 for details).

The final dataset includes 95 misleading images from fact checks collected between August 19 and September 1, 2024, search results for each misleading image, and their unique search result links and snapshots. For each misleading image, we retrieved 7 non-consecutive days of search results over a 15-day window (d1 to d15) following its initial publication date (d0), resulting in a total collection period from August 20 to September 16, 2024 with 34,486 total search results, 7,918 unique search result links and their snapshots, and 7,979 unique pairs of misleading images and search result links. An illustrated glossary of study terms is provided in Figure 2.

**Measures**

***Information Relevance and Veracity***

Three authors manually annotated information relevance and veracity of unique search result links by referencing their original fact-checking articles where the misleading images were identified. Unique search result links appearing under multiple misleading images were annotated separately for each case. A search result link was considered "relevant" if it contained the same misleading image and referred to the exact same issue as the referenced fact-checking article; otherwise, it was considered irrelevant. For example, when an AI-generated image depicts the 9/11 attacks, a search result link that used the same image and addressed topics related to the 9/11 attacks was coded as "relevant."

For search results coded as "relevant", we further coded their veracity into three categories. "Repeated misinformation" restates the false claims recorded by the original fact-checking articles. "Debunking misinformation" refutes the visual misinformation recorded by the fact-checking article. All other search results were coded as "Other relevant information" if they did not directly refute or repeat the original misinformation, but provided relevant information (e.g., a social media post questioning the authenticity of a misleading image without refuting it). Detailed codebooks can be found in Appendix B.1. We conducted three rounds of training and coding, with each round sampling more than 200 unique search result links to resolve discrepancies and refine the codebook, reaching satisfactory intercoder agreement (Krippendorf's $\alpha$ = 0.90, agreement: 95.56% for relevance; Krippendorf's $\alpha$ = 0.84, agreement = 91.67% for veracity). Three coders then coded all unique search result links.

***Search Result Ranking***

The ranking of each search result was derived from its position on the results pages. We assigned the top 10 results ascending ranks from 1 to 10 on "Exact Matches" pages. On "Visual Matches" pages, rankings were assigned to the first 12 results, appearing across the top four rows. Within each row, results were ordered left to right (e.g., positions 1-3 in the first row, 4-6 in the second row).

***Quality of RIS Results Page***

We created a comprehensive score of RIS results page quality, considering search results' information relevance, veracity and their rankings together. Drawing on Hussein et al. (2020), we computed the quality scores of RIS results pages using the following formula:

$$Quality\ Score = \frac{\sum_{r=1}^{n}(x_i * (n - r + 1))}{\frac{n*(n+1)}{2}}$$

where $r$ is the rank of a certain search result and $n$ denotes the total number of search results we obtained from different types of results pages ($n = 10$ for "Visual Matches" pages and 12 for "Exact Matches" pages); $x_i$ equals 10 if a search result debunks misinformation, -10 if it repeats misinformation, and 0 if it is considered "other relevant information" or irrelevant altogether. Thus, the quality score is a continuous value, and on "Exact Matches" pages, it ranges between -10 (e.g., all top 10 results are repeated misinformation) to 10 (all top 10 links are debunking misinformation). Similarly, on "Visual Matches" pages, the value is from -12 to 12. Put simply, a RIS results page has a greater quality score if there are more results with debunking information appearing in more prominent positions, and a lower score otherwise.

### Misleading Image Type and Topic

We manually coded three categories of misleading images: out-of-context, manipulated and AI-generated, according to the original fact-checking articles where the misleading images were sourced. In terms of political versus non-political visual misinformation topics, a misleading image is a political one if its content is related to political figures (e.g., presidents, legislators) or salient political issues (e.g., war, social movement, and key policy debate). Otherwise, the content is non-political. Detailed codebooks can be found in Appendix B.2. Three authors conducted two rounds of coding of all misleading images and resolved the discrepancies, with a Krippendorff's α of 0.78 (agreement = 88.07%) for image type and a Krippendorff's α of 0.77 (agreement = 88.77%) for political topic.

### Search Timing

Search timing was measured as the number of days between the initial publication of the fact-checking article where the misleading image appeared (i.e., d0) and the date of each search to examine temporal variation of RIS results. It ranges from 1 to 15 days.

### Analysis

To answer RQ1, we aggregated all search results and plotted the percentage distribution of content across "Visual Matches" and "Exact Matches" pages, and then conducted one-way ANOVAs comparing content with varied relevance and veracity. To answer RQ2, we plotted the percentage of content at each ranked position to examine variations in relevance and veracity across positions. For RQ3, we plotted the mean quality scores of results pages across misleading image types, political versus non-political topics, and search timing. Also, we conducted linear mixed-effects regressions predicting RIS results page quality scores, with misleading images treated as a random effect and geolocation included as a control variable.

## Results

### Information Relevance, Veracity and Ranking of Search Results

RQ1 asks the distribution of search results regarding relevance and veracity. As shown in Figure 3, on "Visual Matches" pages, there was 78.9% irrelevant content and 10.9% repeated misinformation, with only 7.7% debunking content. On "Exact Matches" pages, the majority of results repeated the misinformation (37.2%), while 32.1% were irrelevant information and 27.7% were debunking content. The ANOVA analyses showed that the observed differences across information relevance and veracity were statistically significant on both "Visual Matches" pages ($F(3, 7,344) = 7,706.93$, $p < .001$), and "Exact Matches" pages ($F(3, 7,304) = 567.251$, $p < .001$). Post-hoc Bonferroni-adjusted pairwise comparisons of different content categories were all significant on both pages (see Appendix C.1).

**Figure 3**

*Percentage of Search Results by Information Relevance and Veracity On "Visual Matches" and "Exact Matches" Pages.*

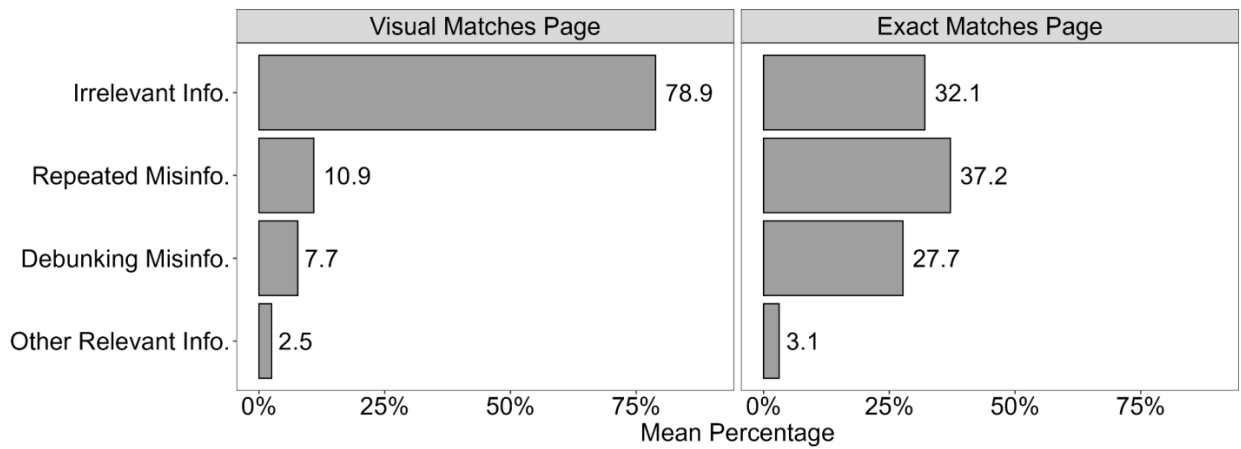

RQ2 asks how search results from RIS are ranked. As shown in Figure 4, on "Visual Matches" pages, across four rows of the collected data, the percentage of irrelevant information decreased with ascending rank within each row, indicating that relevant content tended to appear in the higher-ranked positions. However, the percentage of debunking content (24.5%) exceeded repeated misinformation (16.5 %) only at the top-ranked position (i.e., the first result in the first row) but not at the rest. Notably, the first result of the second row (4th position) which remains highly visible, had the highest likelihood of showing repeated misinformation (31.5 %).

**Figure 4**

*Rankings of Search Results by Information Relevance and Veracity On "Visual Match*es" *and "Exact Match*es" *Pages.*

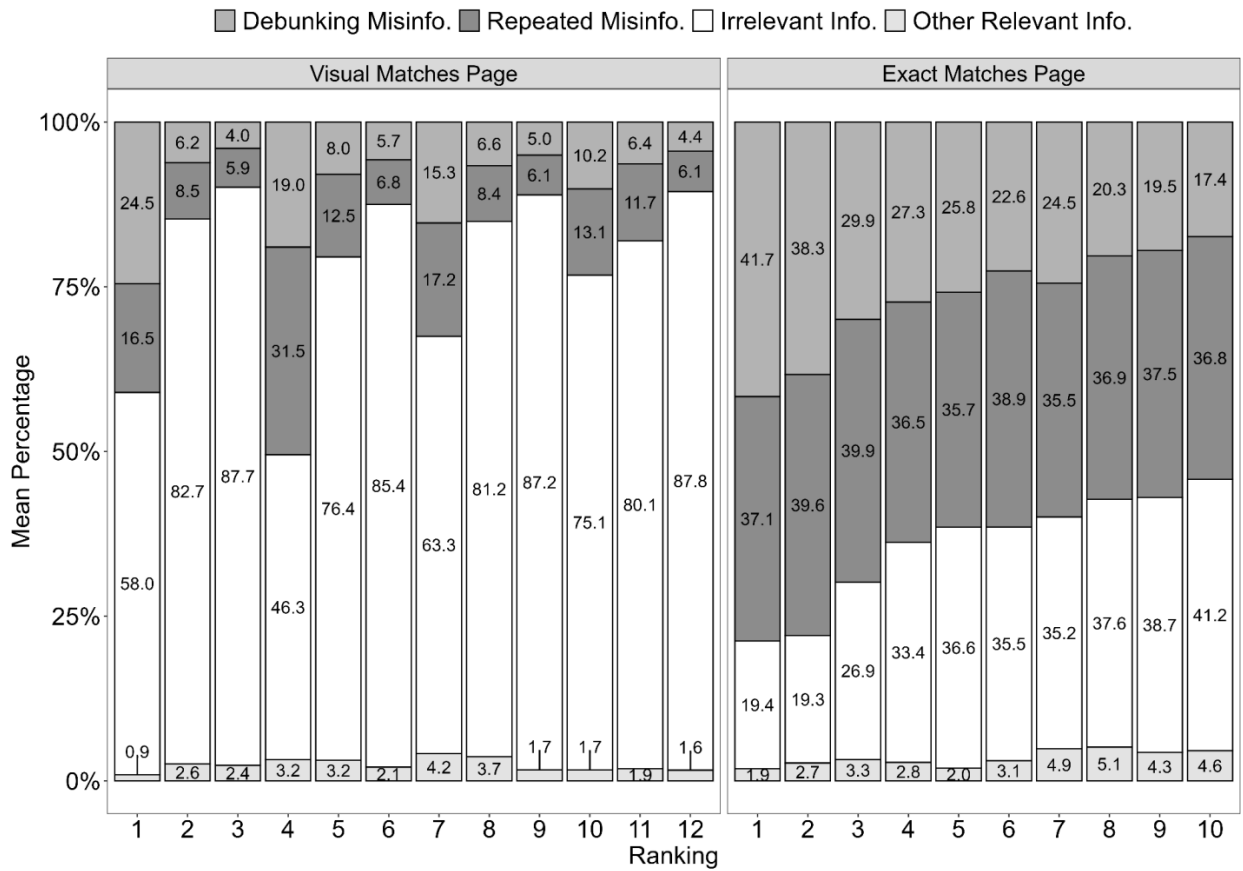

On "Exact Matches" pages, the percentage of debunking content declined from 41.7 % at the top position to 17.4% at the tenth, while irrelevant information increased from 19.4% to 41.2%. The percentage of repeated misinformation remained relatively stable across ranks. Although this pattern suggests that RIS promotes debunking content to higher positions, only the first position displayed a higher proportion of debunking content (41.7%) than repeated misinformation (37.1%), which is consistent with results from "Visual Matches" pages. From the second position onward, users were more likely to encounter repeated misinformation than debunking content.

**RIS Results Page Quality**

RQ3 examines the variations of RIS results page quality. Across all search results, the mean quality scores for "Visual Matches" and "Exact Matches" pages were -0.31 and -0.59 (see Figure 5A), respectively, showing slight negative overall quality.

When comparing three image types, AI-generated misleading images yielded positive mean quality scores on both "Visual Matches" ($M = 3.25$) and "Exact Matches" ($M = 2.17$) pages, whereas out-of-context images were associated with negative quality scores on both ($M = -0.76$ for "Visual Matches" pages; $M = -1.78$ for "Exact Matches" pages). This indicates that RIS may be less reliable in fact-checking out-of-context visual misinformation than AI-generated ones. For manipulated images, the quality scores were positive on "Exact Matches" pages ($M = 0.49$) but negative on "Visual Matches" pages ($M = -0.28$), showing that "Exact Matches" pages could be a relatively better source for verifying manipulated visual misinformation. Regarding image topics, political images can elicit results pages with relatively higher quality scores on both "Visual Matches" and "Exact Matches" pages than non-political ones, though all values remained negative.

Figure 5B illustrates the temporal dynamics, showing that quality scores were relatively stable over time on both pages. However, fluctuations were more pronounced on "Exact Matches" pages, which also exhibited a non-linear pattern with quality score increasing from day 1 to day 5, followed by a subsequent decline through day 15.

**Figure 5**

*Mean RIS Results Page Quality Scores On "Visual Matches" and "Exact Matches" Pages.*

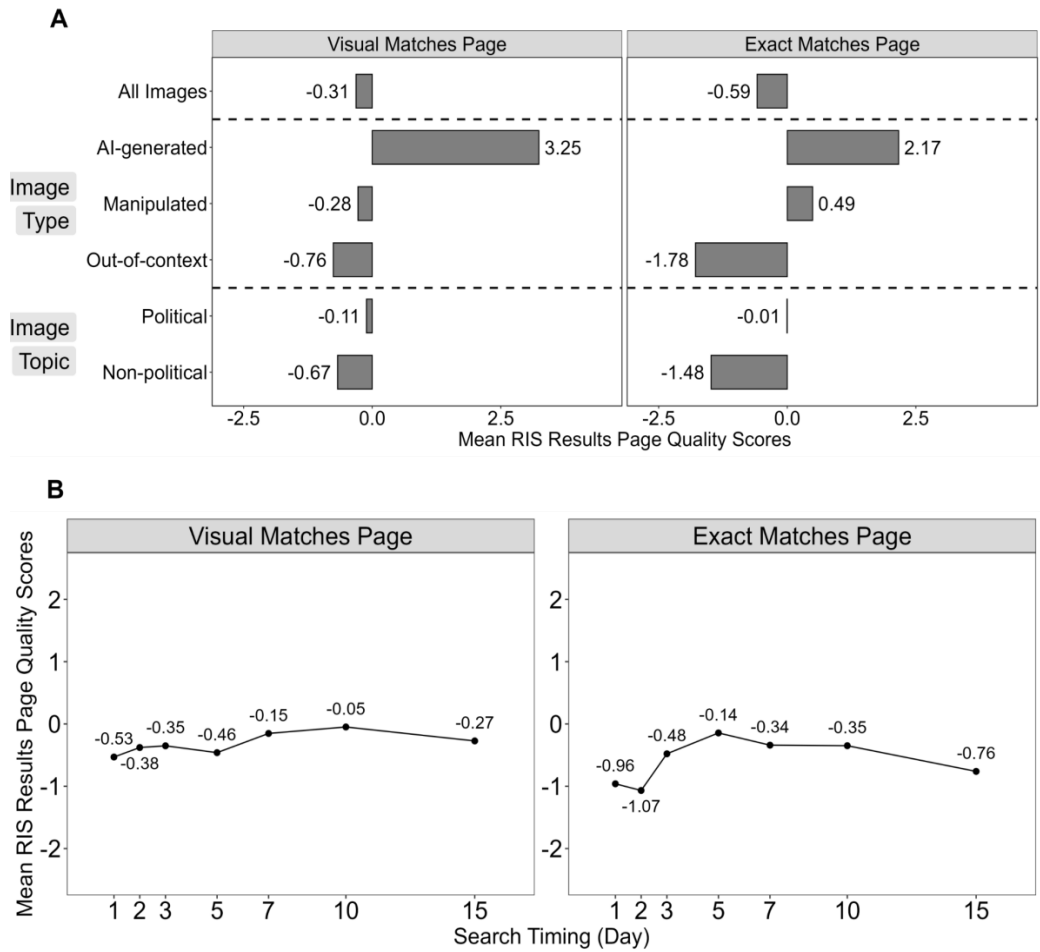

*Note.* Panel A shows mean RIS results page quality scores of all images and images across misleading image types, political versus non-political topics; Panel B shows mean quality scores of all images over time.

The regression model results are shown in Table 1. Compared to out-of-context images,

AI-generated images got higher quality scores on the "Visual Matches" page (*b* = 3.817, S.E. =

0.917, $p < .001$). The linear terms of search timing ("Day" in Table 1) on the quality scores on "Visual Matches" pages ($b = 0.086$, S.E. $= 0.026$, $p < .001$) and "Exact Matches" pages ($b = 0.251$, S.E. $= 0.046$, $p < .001$) were significant and positive. The quadratic terms of search timing ("Day$^2$" in Table 1) were both negative on the quality of "Visual Matches" ($b = -0.004$, S.E. $= 0.002$, $p = .009$) and "Exact Matches" pages ($b = -0.015$, S.E. $= 0.003$, $p < .001$). As Figure 6 shows, these terms jointly suggest an inverted U-shaped relationship between search timing and quality scores on both "Visual Matches" and "Exact Matches" pages: quality score tended to increase at earlier stages, peak at a certain point, and decrease thereafter.

**Table 1**

*Linear Mixed-effects Regression Models Predicting Quality Scores of RIS Results Pages.*

| | Visual Matches Pages | | Exact Matches Pages | |
|---|---|---|---|---|
| | Estimate | S.E. | Estimate | S.E. |
| Intercept | -0.956* | 0.368 | -2.856** | 0.93 |
| Image Type (ref. = Out-of-context) | | | | |
| AI-generated | 3.817*** | 0.917 | 3.65 | 2.412 |
| Manipulated | 0.311 | 0.426 | 2.026 | 1.112 |
| Image Topic (ref. = Non-political) | | | | |
| Political | 0.124 | 0.428 | 0.892 | 1.106 |
| Search Timing | | | | |
| Linear (Day) | 0.086*** | 0.026 | 0.251*** | 0.046 |
| Quadratic (Day$^2$) | -0.004** | 0.002 | -0.015*** | 0.003 |
| Location (ref. = US East) | | | | |
| US Central | 0.008 | 0.074 | -0.027 | 0.13 |
| US West | -0.011 | 0.074 | -0.012 | 0.131 |
| Random effects | | | | |
| $\sigma^2$ | 1.652 | | 5.192 | |
| $\tau_{Images}$ | 3.649 | | 25.598 | |
| ICC | 0.69 | | 0.83 | |
| Observations | 1,837 | | 1827 | |
| Log Likelihood | -3,249.781 | | -4,311.532 | |
| AIC | 6,519.561 | | 8,643.061 | |
| BIC | 6,574.72 | | 8,698.165 | |

**Figure 6**

*Inverted U-shaped Relationship of Search Timing and Predicted Quality Scores of RIS Results Pages.*

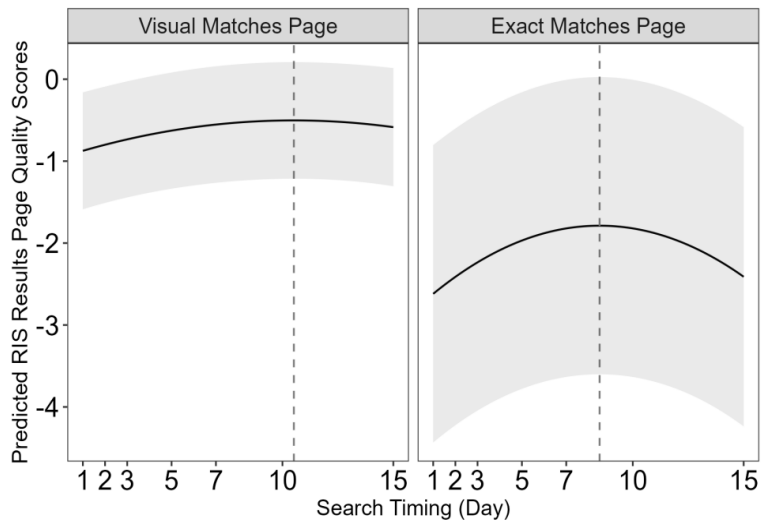

## Discussion

Faced with visual misinformation, individuals often rely on algorithm-driven tools such as search engines to assess information credibility. This study systematically audited the Google RIS, offering a rare empirical evaluation of its gatekeeping role in shaping users' information exposure in self-initiated visual misinformation verification. Although RIS has been widely cited as a promising tool for visual fact-checking, our findings show that, among all search results collected on the first RIS pages, debunking content constitutes less than 10% on "Visual Matches" pages, and less than 30% on "Exact Matches" pages. Meanwhile, more than 80% results on "Visual Matches" pages, and more than 60% on "Exact Matches" pages either repeated false claims or contributed little value for verification, yet they were often placed in prominent positions. We also identified an inverted U-shaped temporal pattern in RIS results page quality, with quality peaking 7-10 days on average after the initial publication of fact-checking articles and declining thereafter. Together, these findings raise concerns about using RIS for visual misinformation verification, and contribute to the gatekeeping theory by extending the algorithmic gatekeeping to visual content.

**Rethinking the Efficacy of RIS for Visual Misinformation Verification**

While existing studies have predominantly explored the text-based misinformation fact-checking (Peter and Koch, 2016; Xue et al., 2024), we argue that an increasingly multimodal environment warrants further exploration of visual misinformation, particularly how widespread algorithmic gatekeepers reshape users' verification outcomes. Consistent with prior research that found text-based search engines return low-quality content that offers limited or even negative utility for fact-checking (Aslett et al., 2024; Hasanain and Elsayed, 2022), our investigation also reveals that Google RIS could potentially amplify visual misinformation, as more than 80% results on "Visual Matches" pages and more than 60% on "Exact Matches" pages are either repeated misinformation or irrelevant information. The potential risk associated with repeated exposure to visual misinformation is well documented: Research suggests that encountering a claim multiple times, regardless of its veracity, can enhance its perceived credibility through processing fluency (Peter and Koch, 2016). When using RIS for verification, users may be inadvertently exposed to visual misinformation repeatedly, potentially reinforcing rather than correcting false beliefs. Also, the high proportion of irrelevant content may increase users' cognitive load, and thus dilute the exposure to debunking content.

The efficacy of RIS is partially dependent on the features of visual misinformation that users try to verify. Out-of-context visual misinformation tends to have lower quality RIS results pages than those of AI-generated visuals. One explanation is that current AI-generated images, especially photorealistic ones, exhibit distinct features such as surrealism and aesthetic professionalism (Peng et al., 2025), which make them easier to identify by human fact-checkers as well as algorithms. Also, fact-checkers are closely monitoring AI-generated falsehoods (Thomson et al., 2024), making fact-checks of AI-generated images more readily available than other visual falsehoods. Moreover, we found no clear difference between RIS results page quality of political and non-political visual misinformation. This likely stems from the narrow operationalization of "political" as election- or politician-related content, while domains such as health or environment also involve public-interest information that is subject to similar algorithmic scrutiny.

**Temporality of RIS Results Page Quality**

Notably, our analysis also demonstrates the temporality of algorithmic gatekeeping in visual misinformation verification: RIS results page quality initially rose with time, peaked around 7-10 days after the initial fact check appeared, and then plateaued or declined. This suggests that, at the very early stages after the publication of fact checks, not much debunking content can be retrieved by RIS. It echoes concerns about "data voids" (Boyd and Golebiewski, 2020), which describes the lack of verification content from credible sources when false claims first emerge. As a result, search engines such as RIS may be exploited by manipulators and low quality websites, thereby surfacing unverified matches and misinformation.

This inverted U-shaped curve also implies that debunking content tends to be most prominent a week to ten days after its publication, but that prominence may not be as durable as one might expect. Prior research shows that, despite consistent debunking of falsehoods, misinformation continues to circulate freely and widely (Singer, 2023). In the current information environment where messages with varied credibility compete for audiences' attention, it is important to consider the constant tug-of-war between organized networks of malicious actors redirecting users towards misinformation (Muñoz et al., 2024) versus the one-shot correction practices of journalists and media organizations. The temporal fluctuation of RIS results page quality may just reflect an eternal battle between content of various veracity, and the publication of fact checks merely marks the beginning, rather than the end, of that battle.

**Algorithmic Gatekeeping in Visual Misinformation Verification**

This study extends the concept of algorithmic gatekeeping to visual content. Compared with text-based searches guided by expertise or trustworthiness principles (Google Search Central, 2025), emerging RIS tools operate primarily through visual embedding and similarity processing (Gaillard and Egyed-Zsigmond, 2017; Hao et al., 2021). Their limited attention to the semantic content and quality criteria means that RIS algorithms may retrieve perceptually similar yet misleading images from unreliable sources.

Also, variation in visual search parameters can therefore result in different RIS outcomes. Our findings illustrate that "Visual Matches" pages, which are designed to retrieve visually *similar* content, contained far less debunking content and far more irrelevant content and misinformation, than "Exact Matches" pages, which prioritize visually *identical* content. This suggests that algorithmic gatekeeping in the visual domain introduces more epistemic uncertainty for misinformation verification: differences between retrieval modes produce substantial shifts in the quality of available evidence, yet these shifts might remain opaque to users relying on visual cues.

From the perspective of user-algorithm interaction, the quality of text-based search results partially depends on users' query inputs, which may or may not align with the false claims being verified. In contrast, images, especially AI-generated or manipulated ones, already constitute falsehoods when submitted into RIS. They bypass the user's query construction process and directly feed misleading visual content into algorithmic systems, thereby facilitating the retrieval of repeated visual misinformation. Also, users process visual information differently from text, as the human cognition privileges visual over textual stimuli (Hockley, 2008). That is, even when textual and visual misinformation appear in comparable positions on search result pages, the visual salience and aesthetic appeal of deceptive images may disproportionately capture user attention and amplify persuasive influence. This highlights the need for developing targeted RIS algorithms to counteract the amplification of visual misinformation, for example, by detecting misalignment between image metadata and accompanying captions, and implementing ranking mechanisms that prioritize authoritative fact-checking sources.

## Implications and Limitations

Methodologically, this study combines the auditing approach with content analysis to examine algorithmic gatekeeping. We extend beyond the one-time collection of algorithmic outputs in previous auditing studies (Juneja and Mitra, 2021; Lin et al., 2023) by repeatedly retrieving RIS results over a 15-day period after the initial fact check of visual misinformation. This design enables a longitudinal and systematic observation of the RIS results over time and can be usefully applied to studying the evolving nature of algorithm-driven ecosystems in other research settings.

Our results also offer practical implications. First, practitioners engaged in fact-checking education or advocacy should be made aware of RIS's limitations and design appropriate training materials and interventions. Meanwhile, fact-checkers should develop targeted strategies addressing distinct challenges posed by different types of visual misinformation (e.g., out-of-context or manipulated images). Second, our study highlights the responsibility of platforms to fine-tune their algorithms to curb the repeated retrieval of misleading images by RIS and enhance accountability in content governance. Finally, this study points out the ways in which users could use RIS more critically, by evaluating features like visual-text alignment and search timing, while navigating the current media environment where truth and falsehood increasingly intertwine.

This study has several limitations. First, the analysis drew from a limited sample of 95 misleading images newly verified by professional fact-checkers in the US. In reality, users situated in different cultural and linguistic contexts often encounter visual misinformation from a more heterogeneous set of sources, many of which are not verified by fact-checkers. Consequently, the scope of this study does not fully capture the complexity of everyday visual misinformation exposure. Second, this study focused solely on top-ranked results on first pages of RIS, and the temporal scope of the data collection was limited to seven non-consecutive days within a 15-day window after the initial publication of fact checks. Expanding search results and extending the data collection period may yield richer and potentially different findings. Still, our focus on top-ranked results within a 15-day window is informed by prior research on users' limited attention and its rapid decay in online information environments (Graffius, 2022; Wu and Huberman, 2007), and is therefore likely to capture the majority of users' experiences. Third, while this study audited the quality of Google RIS results, it did not examine real users' actual exposure and perceptions of misinformation. Future research should extend this work by employing experimental designs, to investigate how users interact with RIS outputs. Also, researchers could further examine how individual cognitive constraints, the attention-grabbing features of visual misinformation, and the dynamic competition between true and false information (Lanham, 2006; Vosoughi et al., 2018) shape users' fact-checking practices via RIS and influence their beliefs in response to misinformation.

### Notes

1. As of 2025, Google occupies the largest market share among various search engines. See https://www.statista.com/statistics/216573/worldwide-market-share-of-search-engines/

2. All data and codes can be found at https://osf.io/f8us9/overview?view_only=f1ee786a1a7b4bac875f357c6a1b72f7.

3. For "Visual Matches", see https://serpapi.com/google-lens-visual-matches-api; for "Exact Matches", see https://serpapi.com/google-lens-exact-matches-api

4. The number of columns may vary due to differences in browser window size or pixel settings. In our experiments, the "Visual Matches" results page was displayed in a three-column layout.

**Supplementary Information**

**From Verification to Amplification? Auditing Reverse Image Search as Algorithmic**

**Gatekeeping in Visual Misinformation Fact-checking**

**Contents**

**Appendix A: More Information on Data Collection**

## A.1. Selection of Visual Misinformation Sources.

We compiled fact-checking websites from the lists of credible fact-checking organizations published by Duke Reporters' Lab (https://reporterslab.org/fact-checking/) and International Fact-Checking Network (https://www.poynter.org/ifcn/). Previous literature has used these lists (e.g., Graves and Mantzarlis, 2020; Liu et al., 2025; Vinhas and Bastos, 2022). We selected fact-checking websites that satisfied all following criteria: 1) it had published fact-checking articles within the past 7 days before data collection started, 2) it is based in the U.S., and 3) it had published image-related fact-checking content in 2024. At the end, 11 fact-checking websites were included.

**Table A.1.** Eleven fact-checking organizations and their URLs.

| Name | URL |
|------|-----|
| AFP Fact Check | https://factcheck.afp.com/ |
| Check Your Fact | https://checkyourfact.com/ |
| FactCheck.org | https://www.factcheck.org/ |
| Lead Stories | https://leadstories.com/ |
| PolitiFact | https://www.politifact.com/ |
| Reuters Fact Check | https://www.reuters.com/fact-check/ |
| Snopes | https://www.snopes.com/latest/ |
| The Dispatch Fact Check | https://thedispatch.com/category/fact-check/ |
| USA Today Fact Check | https://www.usatoday.com/news/factcheck/ |
| Verify This | https://www.verifythis.com/ |
| Washington Post Fact Check | https://www.washingtonpost.com/politics/fact-checker/ |

**A.2. Selection Criteria for Misleading Images**

The selection of visual misinformation was guided by the definition that the visual element misguides audiences through false and inaccurate presentations of information, including manipulating images, images misplaced out of context, and AI-generated images.

Figure A.1 below offers some examples: a1 and a3 were included because certain parts of the two images (e.g., the text) were manipulated to convey misinformation. And a2 was included as it was a totally AI-generated image. For out-of-context visual misinformation, unedited images were retained only when the image itself was contextually specific and constituted a core element of the misleading claim. For example, image a4 was included because the visual features of the image triggered controversy over toilet signage at the 2024 U.S. Democratic National Convention. In contrast, cases in which the misleading claim appeared primarily in the text, while the image merely served as a generic illustration (e.g., general photographs of celebrities or politicians in b), were excluded, as the image did not materially contribute to the misinformation.

**Figure A.1** Examples of included and excluded misleading images.

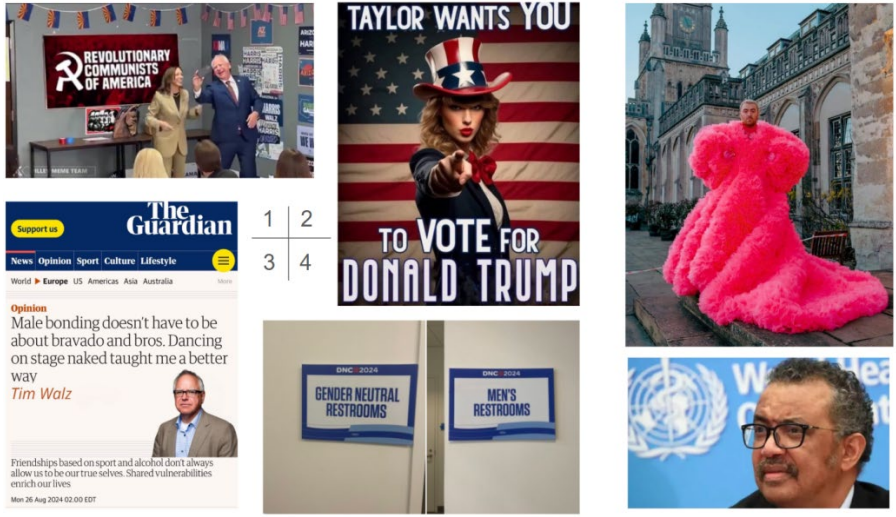

(a) Included visual misinformation          (b) Excluded visual content

**A.3. An Investigation of Repeated RIS Result Links**

A pilot study with 80 unique search result links on three days showed that there were no changes of snapshots of search result links after the initial publication. During the official data collection process, from August 20th, 2024, to September 16th, 2024, we further sampled 10% of all unique links from each day and repeatedly saved the snapshots of a total of 274 search result links on the seven days ($N = 1918$) in a 15-day time span. We found that only 10 (3.6%) of the links underwent changes. For the 10 links, only one social media link was tagged "misleading" starting from the second day and other links cannot be opened on later days during the period. Therefore, we believe downloading the snapshots of repeated links only once would not significantly impact the RIS results.

**Appendix B. Codebooks and Human Coder Training**

**B.1. Information Relevance and Veracity**

The human annotators executed the coding process based on iteratively developed coding schemes. Three rounds of training have been conducted with samples from 200 to 500 in each round. Three coders also discussed their individual annotation heuristics and resolved the discrepancies. To evaluate the reliability of the results, we used Krippendorff's α to assess intercoder agreement. The final intercoder reliability was satisfactory, with Krippendorff's α values of 0.90 for relevance (agreement: 95.56%) and 0.84 for veracity (agreement: 91.67%). The codebook can be found below.

**Table B.1**. Code book for information relevance.

| Category | Definitions | Code |
|----------|-------------|------|
| Relevant | A link is seen as relevant when its content meets the following two criteria:<br><br>   a.   The content uses the same image as our visual misinformation.<br><br>   b.   The content in the link is about the exact same issue and event as in the fact-checking article of the relevant misinformation.<br><br>Example:<br><br>If the fact-checked image is an AI-generated image about 2001, 9/11 attacks, a search result link containing the same image and talking about the related topic is considered "Relevant". | 1 |

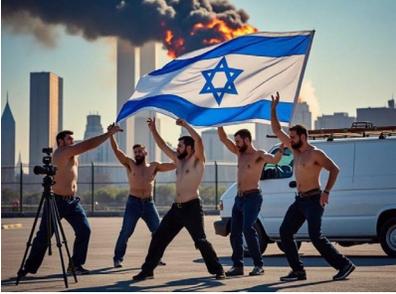

**Fact-checking article**: https://leadstories.com/hoax-alert/2024/08/fact-check-fake-photo-shows-five-dancing-israelis-on-september-11-2001-world-trade-center-attacks.html

**Search result link:**
https://x.com/CensoredMen/status/1823843057869164644

| | | |
|---|---|---|
| Irrelevant | A link is coded as "Irrelevant" if the link content does not contain the misleading image that we are fact-checking. | 0 |

**Table B.2**. Code book for information veracity.

| Category | Definitions | Code |
|---|---|---|
| Repeated misinformation | The link content repeats the relevant false claims that are related to our visual misinformation based on the fact-checking article of the corresponding misleading image.<br><br>Example:<br><br>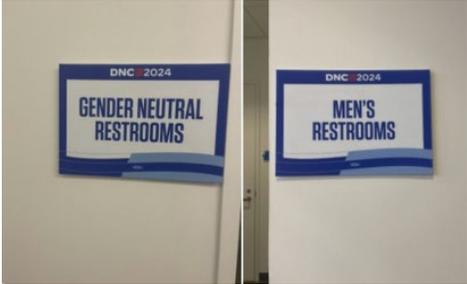<br><br>**Fact-checking article**:  https://leadstories.com/hoax-alert/2024/08/fact-check-dnc-did-not-eliminate-womens-bathrooms-turn-them-into-gender-neutral-bathrooms.html<br><br>**Search result link:**<br>https://x.com/SarahisCensored/status/1825692785565184203 | 1 |
| Debunking misinformation | Content that tags or refutes the visual misinformation, such as:<br><br>a. debunking articles from fact-checking organizations/news media;<br>b. posts that are tagged on the social media platform;<br>c. posts that contain correct information/stating the misinformation is incorrect.<br><br>Example: | 2 |

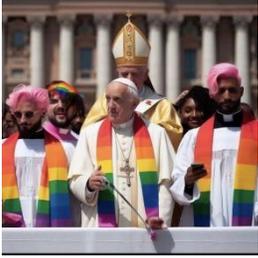

**Fact-checking article**: https://leadstories.com/hoax-alert/2024/08/fact-check-photo-does-not-genuinely-show-pope-francis-wearing-rainbow-scarf.html

**Search result link:** https://x.com/brixwe/status/1824084275257294850

| | | |
|---|---|---|
| Other information | Relevant, but non-misinformation, and non-debunking content, providing original context or questioning cues for the misinformation. For example:<br><br>a. a post that puts forward questions on the visual misinformation;<br>b. the same image published in a different context<br><br>Example:<br><br>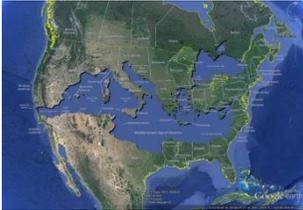<br><br>**Fact-checking article**: https://www.snopes.com/fact-check/us-map-climate-change/<br><br>**Search result link:**<br>https://www.ecoclimax.com/2020/07/mediterranean-sea.html | 3 |

## B.2. Misleading Image Type and Topic

For visual misinformation types, two rounds of training of three human coders were conducted, and the final agreement reached 88.07% with a satisfactory Krippendorff's alpha value (0.778). The codebook can be found below.

**Table B.3.** The codebook of visual misinformation type.

| Category | Definitions | Code |
|---|---|---|
| AI-generated | Fact-checking articles explicitly mentioned that the visual content is wholly fabricated by AI tools.<br><br>Example:<br><br>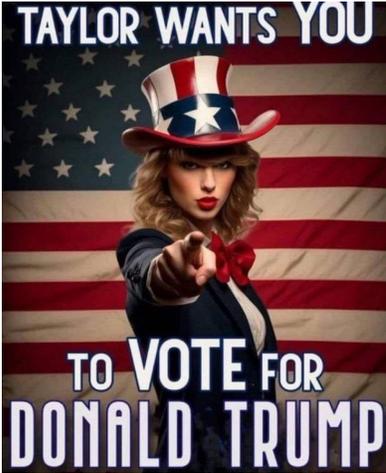<br><br>Fact-checking article: https://leadstories.com/hoax-alert/2024/08/fact-check-taylor-swift-did-not-endorse-donald-trump-in-august-2024.html | 1 |
| | 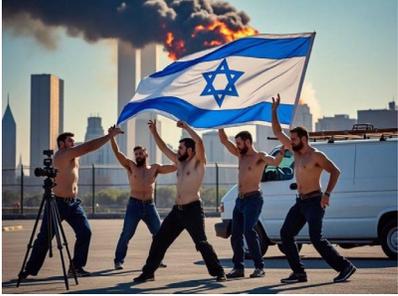<br><br>Fact-checking article:  https://leadstories.com/hoax-alert/2024/08/fact-check-fake-photo-shows-five-dancing-israelis-on-september-11-2001-world-trade-center-attacks.html | |
| Manipulated | Visual content that has been modified in some way to guide the interpretation of the visual information or the larger message unit<br><br>Example:<br>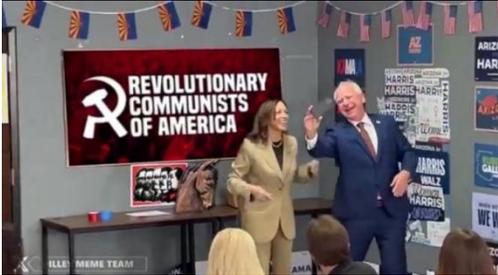<br><br>Fact-checking article:<br>https://factcheck.afp.com/doc.afp.com.36EP9TP | 2 |

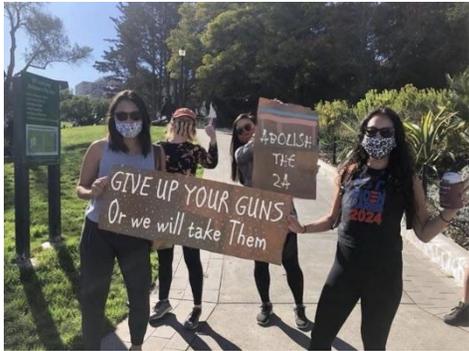

Fact-checking article: https://leadstories.com/hoax-

alert/2024/08/fact-check-fake-photo-shows-biden-harris-supporters-

holding-give-up-your-guns-and-abolish-the-2a-signs.html

| Out-of-context | Unedited and generally authentic visual content alongside text or audio that provides inaccurate or misleading information or context. <br><br> Example: <br><br> 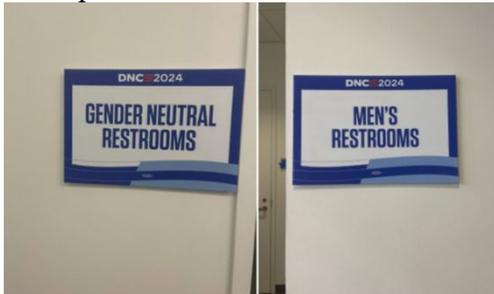 <br><br> Fact-checking article: https://leadstories.com/hoax-alert/2024/08/fact-check-dnc-did-not-eliminate-womens-bathrooms-turn-them-into-gender-neutral-bathrooms.html | 3 |
| --- | --- | --- |

For political versus non-political visual misinformation, two rounds of training of three human coders were conducted, and the final agreement reached 88.77% with a satisfactory Krippendorff's alpha value (0.765). The codebook can be found below.

**Table B.4.** The codebook of visual misinformation topic.

| Category | Definitions | Code |
|----------|-------------|------|
| Non-political | Content that is oriented towards areas including lifestyle, entertainment, sports, and science.<br><br>Example:<br><br>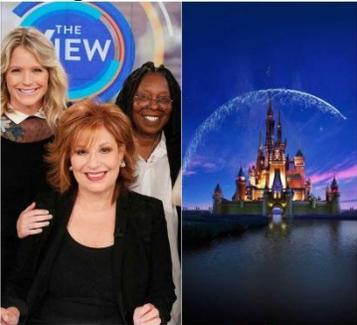<br><br>Fact-checking article: https://checkyourfact.com/2024/08/23/fact-check-disney-selling-abc-20-billion/ | 0 |
| Political | Content related to political figures (e.g., president and legislators) and salient political issues (e.g., war, social movement, and key policy debate)<br><br>Example:<br><br>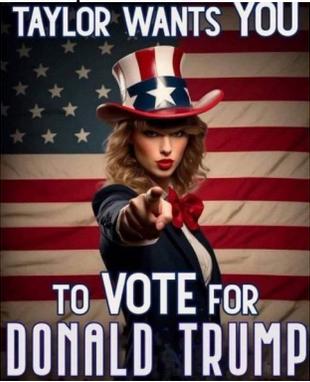<br><br>Fact-checking article: https://leadstories.com/hoax-alert/2024/08/fact-check-taylor-swift-did-not-endorse-donald-trump-in-august-2024.html | 1 |

## Appendix C: Additional Results

## C.1. Post-hoc Pairwise Comparisons

To compare the percentage distributions of content with varied relevance and veracity, we performed one-way ANOVAs. Significant results were found on both "Visual Matches" page ($F(3, 7,344) = 7,706.93$, $p < .001$), and "Exact matches" page ($F(3, 7,304) = 567.251$, $p < .001$). We also conducted Bonferroni-adjusted post-hoc pairwise comparisons of the percentages of different content. The results can be seen in Table C.1, which shows that on both "Visual Matches" and "Exact matches" pages, there is more repeated misinformation than debunking misinformation.

**Table C.1.** Post-hoc pairwise comparisons (Bonferroni-adjusted) of the percentages of content with varied relevance and veracity.

|  | Estimate | S.E. | df | t | p-value | Cohen's d |
|---|---|---|---|---|---|---|
| "Visual Matches" Page |  |  |  |  |  |  |
| Irrelevant Info. - Debunking Misinfo. | 0.711 | 0.006 | 7,344 | 122.404 | <.001 | 4.039 |
| Repeated Misinfo. - Debunking Misinfo. | 0.032 | 0.006 | 7,344 | 5.457 | <.001 | 0.180 |
| Repeated Misinfo. - Irrelevant Info. | -0.680 | 0.006 | 7,344 | -116.947 | <.001 | -3.859 |
| Other Info. - Irrelevant Info. | -0.764 | 0.006 | 7,344 | -131.389 | <.001 | -4.335 |
| Other Info. - Repeated Misinfo. | -0.084 | 0.006 | 7,344 | -14.442 | <.001 | -0.477 |
| Other Info. - Debunking Misinfo. | -0.052 | 0.006 | 7,344 | -8.985 | <.001 | -0.296 |
| "Exact Matches" Page |  |  |  |  |  |  |
| Irrelevant Info. - Debunking Misinfo. | 0.044 | 0.009 | 7,304 | 4.851 | <.001 | 0.161 |
| Repeated Misinfo. - Debunking Misinfo. | 0.094 | 0.009 | 7,304 | 10.52 | <.001 | 0.348 |
| Repeated Misinfo. - Irrelevant Info. | 0.051 | 0.009 | 7,304 | 5.668 | <.001 | 0.188 |
| Other Info. - Irrelevant Info. | -0.290 | 0.009 | 7,304 | -32.294 | <.001 | -1.068 |
| Other Info. - Repeated Misinfo. | -0.341 | 0.009 | 7,304 | -37.963 | <.001 | -1.256 |
| Other Info. - Debunking Misinfo. | -0.246 | 0.009 | 7,304 | -27.443 | <.001 | -0.908 |